\documentclass[prd,aps,showpacs,amsmath,amssymb,nofootinbib]{revtex4}
\usepackage{graphicx}
\usepackage{dcolumn}
\usepackage{bm}

\begin{document}

%

\let\a=\alpha      \let\b=\beta       \let\c=\chi        \let\d=\delta
\let\e=\varepsilon \let\f=\varphi     \let\g=\gamma      \let\h=\eta
\let\k=\kappa      \let\l=\lambda     \let\m=\mu
\let\o=\omega      \let\r=\varrho     \let\s=\sigma
\let\t=\tau        \let\th=\vartheta  \let\y=\upsilon    \let\x=\xi
\let\z=\zeta       \let\io=\iota      \let\vp=\varpi     \let\ro=\rho
\let\ph=\phi       \let\ep=\epsilon   \let\te=\theta
\let\n=\nu
\let\D=\Delta   \let\F=\Phi    \let\G=\Gamma  \let\L=\Lambda
\let\O=\Omega   \let\P=\Pi     \let\Ps=\Psi   \let\Si=\Sigma
\let\Th=\Theta  \let\X=\Xi     \let\Y=\Upsilon

%

%

\def\cA{{\cal A}}                \def\cB{{\cal B}}
\def\cC{{\cal C}}                \def\cD{{\cal D}}
\def\cE{{\cal E}}                \def\cF{{\cal F}}
\def\cG{{\cal G}}                \def\cH{{\cal H}}
\def\cI{{\cal I}}                \def\cJ{{\cal J}}
\def\cK{{\cal K}}                \def\cL{{\cal L}}
\def\cM{{\cal M}}                \def\cN{{\cal N}}
\def\cO{{\cal O}}                \def\cP{{\cal P}}
\def\cQ{{\cal Q}}                \def\cR{{\cal R}}
\def\cS{{\cal S}}                \def\cT{{\cal T}}
\def\cU{{\cal U}}                \def\cV{{\cal V}}
\def\cW{{\cal W}}                \def\cX{{\cal X}}
\def\cY{{\cal Y}}                \def\cZ{{\cal Z}}

\def\dbd{{$0\nu 2\beta\,$}}
%

\newcommand{\Ns}{N\hspace{-4.7mm}\not\hspace{2.7mm}}
\newcommand{\qs}{q\hspace{-3.7mm}\not\hspace{3.4mm}}
\newcommand{\ps}{p\hspace{-3.3mm}\not\hspace{1.2mm}}
\newcommand{\ks}{k\hspace{-3.3mm}\not\hspace{1.2mm}}
\newcommand{\des}{\partial\hspace{-4.mm}\not\hspace{2.5mm}}
\newcommand{\desco}{D\hspace{-4mm}\not\hspace{2mm}}

\newcommand{\beq}{\begin{eqnarray}}
\newcommand{\eeq}{\end{eqnarray}}
\renewcommand\d{\partial}
\newcommand{\tr}{\mathop{\mathrm{tr}}}
 \title{\boldmath $B\rightarrow K^{*}l^+ l^-$: Zeroes of angular observables as test of standard model}

\author{Girish Kumar$^{a,b}$ and Namit Mahajan$^a$
}
\email{girishk@prl.res.in, nmahajan@prl.res.in}
\affiliation{
 {$^a$Theoretical Physics Division, Physical Research Laboratory, Navrangpura, Ahmedabad
380 009, India }\\
{$^b$ Department of Physics, Indian Institute of Technology, Gandhinagar, Ahmedabad 382 424, India}
}

\begin{abstract}
We calculate the zeroes of
angular observables $P_4^{'}$ and $P_5^{'}$ of 4 - body angular distribution of $B\rightarrow K^{*} (\rightarrow K \pi) l^+ l^-$
 where LHCb, in its analysis of form factor independent angular observables, has found deviations from 
 standard model predictions in one of the $q^2$ bins.
 In the large recoil region, we obtain relations between the zeroes of 
 $P_4^{'}$, $P_5^{'}$ and the zero of forward-backward asymmetry of lepton pair.
 These relations, in the considered region, are independent of hadronic uncertainties and depend
 only on Wilson coefficients. We also construct a new observable, 
 $\mathcal{O}_T^{L,R}$, whose zero in the standard model coincides with the zero of forward-backward asymmetry
 but in presence of new physics contributions will show different 
 behavior. Moreover, the profile of the new observable, even within the standard model, is very different from 
 the forward backward asymmetry. We point out that precise measurements of these zeroes in near future
 would provide crucial test of the standard model and 
 would be useful in distinguishing between different possible new physics contributions to Wilson coefficients. 
 %
%
\end{abstract}

\maketitle

\section{Introduction}

Rare B decays are mediated by flavor changing neutral current (FCNC) $ b \rightarrow s$ transitions which are absent in 
the standard model (SM) at tree level. 
The leading contribution comes from one loop diagrams. Being GIM and CKM suppressed, their predictions in SM are very tiny. 
As these processes are very
sensitive to heavy particles in the loops, any effect from new physics (NP) will show significant deviation from SM predictions. 
This make these decays
assets for indirect probes of NP. So far data collected by dedicated experiments (LHCb, B-factories) on rare B-decays 
 are in excellent agreement with the predictions of SM and have been used to retrieve information on flavor structure
of possible new physics and have put stringent constraints on beyond Standard Model (BSM) scenarios 
but expectations of looking for any definitive hints of NP have not met with success. The results seems to be in consistency 
with CKM mechanism of SM \cite{Kobayashi:1973fv}. However, recent data on angular observables of
$B\rightarrow K^{*}(\rightarrow K \pi) l^+ l^-$ four body distribution 
indicate a plausible change in this situation. LHCb has observed a 3.7$\sigma$ level deviation from SM prediction in 
one of the angular observables, $P_5^{'}$ \cite{Aaij:2013qta}. 
This discrepancy might be a result of statistical 
fluctuations or inevitable theoretical uncertaiities that enter in calculation of these observables \cite{Das:2012kz}.
One has to wait for more experimental 
data and a more careful analysis of theoretical uncertainties to clear the smoke. 
Assuming that this discrepency is solely due to NP effects,
there have been attempts in literature to resolve this tension between theory and experimental side with 
solutions having various NP scenarios
(see for example \cite{Descotes-Genon:2013wba}).  \\
In this piece of work, we study some of the angular observables $P_4^{'}$, $P_5^{'}$, $A_{FB}$ and a new observable, 
which we we call $\mathcal{O}_T^{L,R}$, 
with a different  approach. We look at zeroes of these observables. The expressions, under certain reasonable assumptions,  
are more or less independent 
of theoretical  uncertainties, and depend solely on short distance Wilson coefficients,
and thus have very clean predictions in SM. Precise measurement of 
 these quantities gives certain  relations among Wilson coefficients and thus provide very 'clean test' of SM.\\
We proceed as follows. In the next section, we define effective Hamiltonian for $b\rightarrow s l^+ l^-$. In section III, 
we discuss the 4-body angular
distribution of $B\rightarrow K^{*}(\rightarrow K \pi) l^+ l^-$ and various observables in in large energy recoil limit.
In section IV, we calculate zeroes of these observables including a new observable 
$\mathcal{O}_T^{L,R}$ and obtain correlation among them. In section V, we summarise and conclude.

\section{Effective Hamiltonian}
 The basic framework to study rare FCNC decays where QCD corrections can give sizeable contributions is that
 of effective Hamiltonian which 
 is obtained after integrating out the heavy degrees of freedom. 
 The rare decay $B \rightarrow K^* l^+ l^-$ is governed by effective Hamiltonian for $ b\rightarrow s $ transitions written as  
\begin{equation}
 \mathcal{H}_{eff} = -\frac{4 G_F}{\sqrt{2}} V_{ts}^{*} V_{tb} \sum_i (C_i (\mu)O_i + C_i^{'}(\mu) O_i^{'}) + \mbox{h.c.}
\end{equation}
where contribution of the term $\propto \frac{V_{ub} V_{us}^{*}}{V_{tb} V_{ts}^{*}}$ is neglected. 
$O_i$ are the effective
local operators and their coefficients $C_i(\mu)$ are called Wilson coefficients evaluated at scale $\mu$. 
The factorization scale $\mu$ distinguishes between
short distance physics (above scale $\mu$) and long distance physics (below scale $\mu$).
Wilson coefficients depend on factorization scale and are the
only source of information about heavy degrees of freedom which have been integrated out while 
matrix elements of local operators $O_i$
dictate the low energy dynamics (for a review see \cite{Buchalla:1995vs}). 

The operators contributing significantly to the process $B \rightarrow K^* l^+ l^-$ in SM are semileptonic vector operator 
$O_9$, axial vector operator 
$O_{10}$ and magnetic photon penguin operator $O_7$. Their explicit form is given by
\begin{equation}
 \begin{split}
 O_7 &= \frac{e}{16 \pi^2} m_b (\bar{s}_\alpha \sigma_{\mu\nu} R b_\alpha) F^{\mu\nu}, \\
 O_9 &= \frac{e^2}{16 \pi^2}(\bar{s}_\alpha\gamma^\mu L b_\alpha)(\bar{l} \gamma_\mu l),\\
 O_{10} &= \frac{e^2}{16 \pi^2}(\bar{s}_\alpha\gamma^\mu L b_\alpha)(\bar{l} \gamma_\mu\gamma_5 l),
 \end{split}
\end{equation}
Here, $\alpha,\,\,\beta \,$ are the color indices,  $L,R = \frac{(1\mp\gamma_5)}{2}$ represent chiral projections, 
$T^a$ are the  SU(3) color charges. 
$m_b$ is the b-quark mass.
The  primed operators have same tensorial structure as unprimed ones
but with helicity flipped. Their contribution within SM is severely suppressed or vanishes.\\ 

 The 
 effective coefficient  of operator $O_9$ is given by 
\begin{equation}
 C_9^{eff} = C_9 + Y(\hat{s})
\end{equation}
Here $s$ is lepton invariant mass and $\hat{s}$ is the  invariant mass $(s)$ normalized by B-meson mass square,
i.e., $\hat{s} = s/m_B^2$.  $ Y(\hat{s})$  
is one loop function and  contains contribution from one loop matrix elements of operators $O_{1,2,3,4,5,6}$.  
The form of function $Y(\hat{s})$ can be found in \cite{Asatrian:2002va}. Due to $Y(\hat{s})$,
$C_9^{eff}$ is not real but has a small imaginary part. 
For calculating zeroes of different observables and getting analytic relations among them, we will
treat these Wilson coefficients as real and also neglect small $Y(\hat{s})$ but 
for numerical calculations we include $Y(\hat{s})$ in $C_9^{eff}$. As will be evident later, this turns out to be
a good working approximation.

\section{Angular Observables of $B\rightarrow K^* l^+l^-$ in large recoil limit}

 To calculate observables for $B\rightarrow K^*$ process, one needs to calculate matrix elements of the local operators $O_is$. 
These matrix elements are usually expressed in terms of seven form factors $V, A_0, A_1, A_2, T_1,T_2$ and $T_3$ which 
are functions of momentum
transfer between B and $K^*$.
These form factors are calculated via non-perturbative methods like QCD sum rules on the light cone (LCSRs) \cite{Ball:2004rg} 
when daughter mesons energies are large. Working in QCD factorization (QCDF) framework, and within
heavy quark and large recoil limit, 
all seven seven form factors can be written in terms of 
only two independent universal factors, namely, $\xi_\bot$ and $\xi_\parallel$ \cite{Charles:1998dr}. 
The decay amplitude can be represented as
$\sim C_a \xi_a + \Phi_B \otimes T_a \otimes \Phi_{K^{*}}$, with 'a' corresponding to polarization of $K^{*}$,
$\bot, \parallel$. $C_a$ contains 
factorizable and non factorizable correction which are calculated
in perturbation theory with the help of renormalization group (RG) techniques.\\
The two set of form factors are related to each other through following identities (see for example \cite{Beneke:2004dp})
\begin{equation}
\begin{split}
 \xi_\bot &= \frac{m_B}{m_B + m_K^*} V(q^2),\\
 \xi_\parallel &= \frac{m_B + m_K^*}{2 E} A_1(q^2) - \frac{m_B - m_K^*}{m_B} A_2(q^2)
\end{split}
\end{equation}

 
 The 4-body angular distribution of $b\rightarrow K^*(\rightarrow K \pi) l^+ l^-$ offers experimentally accessible observables
which are independent of form factors and hence theoretically cleaner. The fully differential decay distribution can be described 
by four kinematical variables, given by
\begin{equation}
 \frac{d^4\Gamma(b\rightarrow K^*(\rightarrow K \pi) l^+ l^-)}{dq^2 d cos\theta_K cos\theta_l d\phi} = 
\frac{9}{32 \pi} J(q^2,\theta_l,\theta_K,\phi)
\end{equation}
where kinematical variables dilepton invariant mass $q^2$, $\theta_l$, $\theta_K$ and $\phi$ are defined in 
\cite{Kruger:1999xa} and 
\begin{equation}
  J(q^2,\theta_l,\theta_K,\phi) = \sum_i J_i(q^2) f(\theta_l,\theta_K,\phi)
\end{equation}
The angular coefficients $J_i(q^2)$ are 
 generally expressed in terms of transversity amplitude. There are in total seven transversity amplitudes. 
 There will be an additional amplitudes 
once scalar interactions are also taken into account which we do not consider in this work. 
At the leading order in $1/m_b$ and 
$\alpha_s$ the transversity amplitudes render to following simple expression:
\begin{equation}
 A_\bot^{L,R} = \sqrt{2} N m_B (1-\hat{s}) \left[(C_9^{eff} + C_9^{' eff}) \mp (C_{10} + C_{10}^{'}) + 
 2\frac{ \hat{m}_b}{\hat{s}} (C_7^{eff} + C_7^{' eff})
 \right] \xi_\bot (E_{K^*}),
\end{equation}

\begin{equation}
 A_\parallel^{L,R} = -\sqrt{2} N m_B (1-\hat{s}) \left[(C_9^{eff} - C_9^{' eff}) \mp (C_{10} - C_{10}^{'}) + 
 2\frac{ \hat{m}_b}{\hat{s}} (C_7^{eff} - C_7^{' eff})
 \right] \xi_\bot (E_{K^*}),
\end{equation}

\begin{equation}
 A_0^{L,R} = -\frac{N m_b}{2 \hat{m}_{K^*}\sqrt{\hat{s}}} (1-\hat{s})^2 \left[(C_9^{eff} - C_9^{' eff})
 \mp (C_{10} - C_{10}^{'}) + 2 \hat{m}_b (C_7^{eff} - 
 C_7^{' eff})\right] \xi_\parallel (E_{K^*}),
\end{equation}

\begin{equation}
 A_t = \frac{N m_b}{\hat{m}_{K^*}\sqrt{\hat{s}}} (1-\hat{s})^2 \left[C_{10} - C_{10}^{'}\right] \xi_\parallel (E_{K^*})
\end{equation}
Here, $\hat{m}_b = m_b/m_B$ and $E_{K^*}$ is the energy of $K^*$ meson. 
Terms of $\mathcal{O}(\hat{m}_{K^{*}}^2)$ have been neglected. However 
It is worth mentioning that these relations holds only in the kinematical region $1 < q^2 < 6$.
This is precisely the region of interest\\

 There are  12 angular coefficients $J_i(q^2)$ and considering $\bar{J}_i(q^2)$ (corresponding to CP conjugate mode of 
 $B \rightarrow K^*(\rightarrow K \pi) l^+ l^-$), there are in total 24 angular coefficients.
 The CP conjugated coefficients $\bar{J}_i$ are given by $J_i$ with 
 weak phases conjugated. One can construct, taking certain ratios and combinations such that form factors and 
 hadronic uncertainties pertaining to such
 observables more or less cancel and we are left with observables which are cleaner and have high sensitivity to NP effect. 
 One can define such CP -averaged 
 and CP violating observables as in \cite{Kruger:1999xa},\cite{Kruger:2005ep}.

\section{Zeroes of angular observables}

It is well known that the zero of the forward backward asymmetry of the lepton pair in the decay
$B\rightarrow K^* l^+ l^-$ is highly insensitive to form factors and precise measurement 
of this quantity can reveal new physics \cite{Burdman:1998mk}.
The zero of forward backward asymmetry is known to depend on ratios of form factors,
value of b quark
mass and Wilson coefficients $C_7^{eff}$ and $C_9^{eff}$. 
However, in the heavy quark limit and in the region where energy of $K^{*}$ $(\sim m_B/2)$ is comparable
with B-meson mass, the hadronic uncertainties cancel in ratios of form factors and, to a good approximation, 
zero of the forward backward asymmetry of the lepton pair is 
essentially independent of form factor uncertainties. The position of the zero is thus heralded as a test of SM since
the position shifts significantly for most models beyond SM.
The zero of the forward backward asymmetry is given by a clean relation:
\begin{equation}
 \mbox{Re}(C_9^{eff}(\hat{s}_0)) = -2 \frac{\hat{m}_b}{\hat{s}_0} C_7^{eff} \frac{1-\hat{s}_0}{1 + \hat{m}_{K^*}^2 -\hat{s}_0}
  \sim -2 \frac{\hat{m}_b}{\hat{s}_0} C_7^{eff}
\end{equation}
Here, $\hat{s}_0$ is position of the zero of the forward-backward asymmetry. 
Taking it as cue, we investigate other observables, particularly $P_5^{'}$ and $P_4^{'}$ as there has been a 
tension between SM prediction  and experimental data for these observables in one of the bins.\\

To bring out the power to differentiate various NP scenarios, we calculate the zeroes of some of the angular
observables including the primed operators as well. The associated Wilson coefficients are assumed to be real for 
simplicity though it is straight forward to generalise the relations below to complex coefficients. This is not
necessary for the present as our main motivation is to study the situation within SM, where as we show below, there are
tight correlations of the zeroes of the angular observables considered and the zero of the forward backward asymmetry.
 We also propose a new observable, $\mathcal{O}_T^{L,R}$, defined below. This new observable, and its zero, carries
 quite a complimentary information compared to the observables already studied in literature.
 To the best of our knowledge, such correlations and their impact as in providing a litmus test for SM
 has not been studied before. We would again like to emphasize that the analytic expressions below have been
 obtained by neglecting the $Y(\hat{s})$ contribution from $C_9^{eff}$ and treating the leptons as massless. However,
 in the numerical evaluations, we have retained the $Y(\hat{s})$ contribution and have massive leptons.\\
 
  To calculate zeros of any of the  observables, one needs to look  solution of numerator only. 
  We set them to zero and obtain the solution.
 \begin{enumerate}
  \item [(a)] $A_{FB}$:
  In terms of the
 angular  coefficients, forward-backward asymmetry is proportional to $(J_{6s} + \bar{J}_{6s})$ which in turn 
 is $\propto 
 [\mbox{Re}(A_\parallel^L A_\bot^{L *}) -  (L\leftrightarrow R)]$.  Setting $(J_{6s} + \bar{J}_{6s}) = 0$ gives following solution
 \begin{equation}
   \hat{s}_0 = -2 \,\frac{( C_{10} C_7 - C_{10}^{'} C_7^{'})}{( C_{10} C_9 - C_{10}^{'} C_9^{'})} \hat{m}_b 
 \end{equation}
\end{enumerate}
\begin{enumerate}
\item [(b)] $P_5^{'}$:
$P_5^{'}$ is proportional to $(J_{5} + \bar{J}_{5})$ which in massless lepton limit is 
$ \propto [\mbox{Re}(A_0^L A_\bot^{L *}) -  (L\leftrightarrow R)] $ . The zero of $P_5^{'}$ is given by 
\begin{equation}
  \hat{s}^{P_5}_0 =  \frac{( C_7 + C_7^{'}) (C_{10}^{'} - C_{10}) }
  {[C_{10} C_9 - C_{10}^{'} C_9^{'} + (C_7-C_7^{'}) (C_{10} + C_{10}^{'}) 
  \hat{m}_b]} \hat{m}_b  
  \end{equation}
\end{enumerate}

\begin{enumerate}
\item [(c)] $P_4^{'}$:
The numerator of $P_4^{'}$ expression is $(J_{4} + \bar{J}_{4}) \propto 
[\mbox{Re}(A_0^L A_\parallel^{L *}) +  (L\leftrightarrow R)]$. The zero of $P_4^{'}$ is given by
\begin{equation}
 \hat{s}^{P_4}_0 =  -2 \frac{( C_7 + C_7^{'})[( C_9 + C_9^{'}) + 4 (C_7 -C_7^{'})] \hat{m}_b}{[(C_9-C_9^{'})^2 +
 (C_{10}-C_{10}^{'})^2 + 2 (C_7-C_7^{'}) 
 (C_9 -C_9^{'}) \hat{m}_b]} \hat{m}_b 
\end{equation}
\end{enumerate}

\begin{enumerate}
\item [(d)] $\mathcal{O}_T^{L,R}$: Apart from these observables, we also construct a new observable,
$\mathcal{O}_T^{L,R}$, which has following form 
\begin{equation}
 \mathcal{O}_T^{L,R} = \frac{|A_{\bot}^L|^2 + |A_{\parallel}^L|^2 -  \,(L \leftrightarrow R)}{\sqrt{-(J_{2s} + 
 \bar{J}_{2s}) (J_{2c} + \bar{J}_{2c})}}
\end{equation}
The zero of the observable $\mathcal{O}_T^{L,R}$ is given by 
\begin{equation}
 \hat{s}^{\mathcal{O}_T^{L,R}}_0 =  -2 \,\frac{( C_{10} C_7 + C_{10}^{'} C_7^{'})}{( C_{10} C_9 + C_{10}^{'} C_9^{'})} \hat{m}_b 
\end{equation}
\end{enumerate}

\subsection*{Relations in Standard Model}
It is very interesting to consider these relations in the limit of SM: set  $C_i^{'} = 0$. Further, we exploit the fact that
within SM, numerically, $C_9 \approx -C_{10}$. Employing these and simplifying, we obtain:
\begin{equation}
 \hat{s}_0^{SM} = -2 \,\frac{C_7}{ C_9 } \hat{m}_b 
\end{equation}
which matches with the relation (11) within large recoil limit. 
The LHCb collaboration has measured the point of zero crossing of the forward backward asymmetry zero:
$q_0^{2}$ = (4.9 $\pm$ 0.9)
$\mbox{GeV}^{2}$ \cite{Aaij:2013iag} \\
In the case of $P_5^{'}$, setting primed Wilson coefficients equal to zero in equation (12), relation reduces to 
\begin{equation}
  \hat{s}^{P_5, SM}_0 = -\frac{ C_7}{ C_9 +C_7\, \hat{m}_b} \hat{m}_b  \frac{\hat{s}_0/2}{1-\hat{s}_0/2} \,\,\approx 
  \frac{\hat{s}_0}{2}
\end{equation}
So the relations predicts value of zero of $P_5^{'}$ to be approximately half of forward-backward
asymmetry zero.  \\

In the SM limit the zero of $P_4^{'}$ reads
\begin{equation}
 \hat{s}^{P_4, SM}_0 =  -2 \frac{C_7  C_9  + 2 C_7^2  \hat{m}_b}{C_{10}^2 + C_9^2 + 2 C_7 C_9 \hat{m}_b} \hat{m}_b 
 \frac{\hat{s}_0(1 - \hat{s}_0)}{(2-\hat{s}_0)}  \,\,\approx  \frac{\hat{s}_0}{2}
\end{equation}
where in the last step, we have again used the fact that $\hat{s}_0$ is very small compared to 1. 
So, the prediction for the zero of $P_4^{'}$  is also about half of $\hat{s}_0$.\\ 

If we keep the effect of  factor $\hat{s}_0$ in expressions of zeroes of $P_5^{'}$ and $P_4^{'}$,
we see that actual value of the zero of $P_5$ is a little larger 
than half of $\hat{s}_0$ while the zero of $P_4^{'}$ lies a bit below half of $\hat{s}_0$,
but as clearly evident from Table 2, the effect is rather small and can be safely neglected. \\

The case of the proposed new observable $\mathcal{O}_T^{L,R}$ is special one. 
The expressions of zero of forward backward asymmetry (12) and $\mathcal{O}_T^{L,R}$ (15) 
have interesting features. In the SM limit, the position of the two zeroes coincides while this degeneracy is
lifted in a simple but complimentary manner when the helicity flipped operators are present.\\


The expressions of zeroes of these observables depend only on Wilson coefficients, practically independent
of form factors, thereby leading to theoretically clean predictions. 
To calculate these zeroes, 
we use $C_9$ = 4.2297, $C_{10}$ = -4.2068, $C_7^{eff}$ = -0.2974 \cite{Hurth:2013ssa} at scale $m_b$. 
In our numerical analysis, we use the values of 
input parameters given in table I. 
\begin{table}
  \caption{Values of input parameters used for numerical calculations of zeroes of observables} 
   \begin{tabular}{ | c c  |}
      \hline\hline
     $m_b^{pole}$ & 4.80 $\mbox{GeV}$\\
     $G_F$ & 1.166$\times 10^{-5}$\\
     $m_B$ & 5.280 $\mbox{GeV}$\\
     $m_{K^{*}}$ & 0.895 $\mbox{Gev}$\\
     $m_\mu$ & 0.106 $\mbox{GeV}$\\
     $\alpha$ & 1/129\\
     $\alpha_s$ & 0.21\\
     \hline
     \hline
   \end{tabular}
\end{table}

 \begin{figure}[ht!]
\vskip 0.32cm
\hskip 1.35cm
\hbox{\hspace{0.03cm}
\hbox{\includegraphics[scale=0.65]{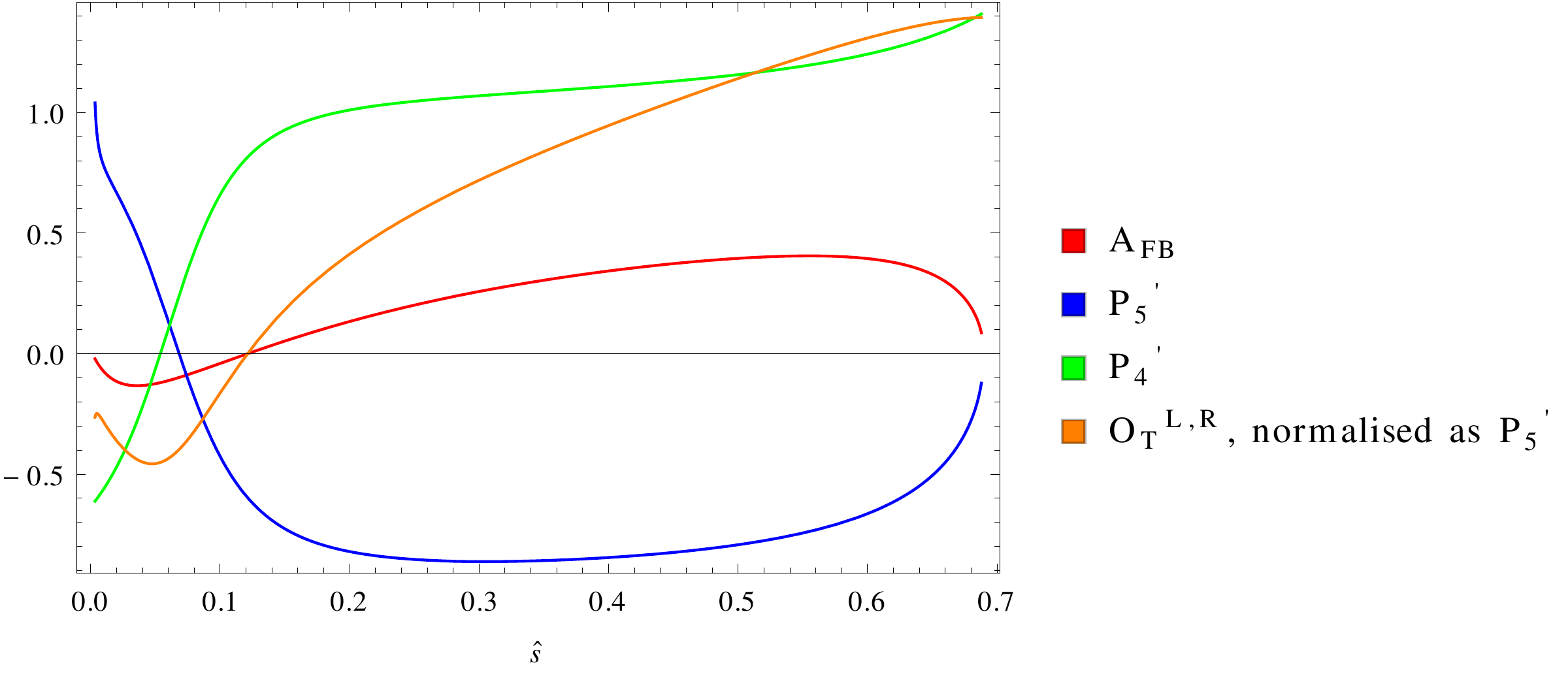}}
}
\caption{Different angular observables as a function of $\hat{s}$
 }
 \label{fig1}
\end{figure}

To compare with the exact predictions in SM and to have a consistency check of these relations, we also calculated  values of zeroes in SM with complete set of
seven form factors and keeping the $Y(\hat{s})$ in $C_9^{eff}$. For numerical calculation,
we use the form factors calculated in \cite{Ball:2004rg} using light-cone sum rule. We tabulate results for zeroes from both 
the approaches 
in table II. It is clear that the employed analytic relations yield values close to those when no approximations are made.
This shows robust nature of these relations.
 As a result, any NP effects present in $C_7$, $C_9$ and $C_{10}$ will be  tightly constrained by the relations (13,15,18,16,19).
 It is found (see \cite{Descotes-Genon:2013wba}) that to explain $P_5^{'}$ anomaly from the perspective of new physics, the new physics has a destructive 
 contribution to $C_9$. 
 The magnitude of $C_7^{eff}$ 
 is very accurately known from B decay $B\rightarrow K^{*} \gamma$. Assuming real coefficients, this then 
 means that $C_7^{eff}$ is known up to a sign ambiguity. This ambiguity in the sign of $C_7$
 is resolved by the zero of the
 forward backward asymmetry of the lepton pair, and therefore precise deduction of Wilson coefficient $C_9$ can be done.
 We would be able to identify distinctions among different NP scenarios more accurately once these zeroes are
 precisely measured. We would also like to draw attention to the fact that not just the position of the zero of
 an angular observable but also the complete profile as a function of $\hat{s}$ is a powerful tool at hand.
 This is illustrated in Figure 1 where one can clearly see that though the zero of forward backward asymmetry
 coincides with that of the new observable $\mathcal{O}_T^{L,R}$, the two profiles are quite different.

 \begin{table}
  \caption{Comparison of values of zeroes of different observables in SM. In column I, the values are calculated 
  using relations (13,15,18,16,19) 
  we obtained  while  the last column has entries predicted by SM with full form factors.} 
   \begin{tabular}{ | c | c | c |}
      \hline
     Observable & Value of zero of observable using relations & Exact value of zero \\ \hline
     $A_{FB}$ & 0.128  & 0.122 \\ \hline
     $P_5^{'}$ & 0.068 & 0.069 \\ \hline
      $P_4^{'}$ &0.059 & 0.055 \\ \hline
     $\mathcal{O}$ &0.128 & 0.122\\
     \hline
   \end{tabular}
\end{table}

\section{Summary and Conclusions}
With the latest experimental results on angular observables of $B\rightarrow K^{*}(\rightarrow K \pi) l^+ l^-$ showing 
discrepancies with respect to SM, one would like to hope
to have found first evidence of new physics. But due to uncertainties inherent in the theoretical calculations of 
such processes, it is difficult to infer the same in affirmation. Precise measurements of theoretically
clean observables holds the best chance of unambiguously revealing the presence of physics
beyond SM, if any. The zero of the forward backward asymmetry is known to fall under this category of observables.
The current measurement is not precise enough to say anything definitive and is totally consistent with SM.
It may be useful to have more such observables measured with precision.
With this picture in mind, we point out that zeroes of forward backward asymmetry, $P_5^{'}$, $P_4^{'}$ and 
$\mathcal{O}_T^{L,R}$ (a new angular observable proposed in this work), can be crucial test of SM. It has been pointed out that
within SM, the position of all the zeroes is essentially fixed by the zero of the forward backward asymmetry, 
up to small corrections. To the best of our knowledge, this is the first attempt to use such correlations as a stringent
test of SM itself. 
The relations are quite rich and general as 
they include Wilson coefficients of 
helicity flipped operators also. The relations are obtained in large recoil region in large energy limit where 
estimates of theoretical uncertainties are supposed to be 
minimalistic. A simultaneous accurate determination of these zeroes will surely provide a conclusive
evidence of any NP present. Moreover, in a general setting, the zeroes by themselves carry complimentary
information about the Wilson coefficients and their measurement together with the existing data can be used to pin point the class
of NP scenarios which can give rise to such predictions. This is clearly evident from the position of the zero of the proposed
observable $\mathcal{O}_T^{L,R}$ which in the standard model limit yields the same value as the zero of forward backward asymmetry
but when the helicity flipped operators are included, leads to complimentary information on the Wilson
coefficients compared to the zero of forward backward asymmetry. We also hope that with more data, not just the position of various
zeroes, but also the complete profiles of angular observables will be known with high precision, which can be used further
as a crucial test of the standard model.


\begin{thebibliography}{99}


\bibitem{Kobayashi:1973fv} 
  M.~Kobayashi and T.~Maskawa,
  Prog.\ Theor.\ Phys.\  {\bf 49}, 652 (1973).



\bibitem{Aaij:2013qta} 
  R.~Aaij {\it et al.}  [LHCb Collaboration],
  Phys.\ Rev.\ Lett.\  {\bf 111}, no. 19, 191801 (2013)
  [arXiv:1308.1707 [hep-ex]].


\bibitem{Das:2012kz} 
  D.~Das and R.~Sinha,
  Phys.\ Rev.\ D {\bf 86}, 056006 (2012)
  [arXiv:1205.1438 [hep-ph]];
  S.~Jäger and J.~Martin Camalich,
  JHEP {\bf 1305}, 043 (2013)
  [arXiv:1212.2263 [hep-ph]];
  C.~Hambrock, G.~Hiller, S.~Schacht and R.~Zwicky,
  Phys.\ Rev.\ D {\bf 89}, 074014 (2014)
  [arXiv:1308.4379 [hep-ph]];
  J.~Lyon and R.~Zwicky,
  arXiv:1406.0566 [hep-ph];
  S.~Descotes-Genon, L.~Hofer, J.~Matias and J.~Virto,
  arXiv:1407.8526 [hep-ph].
  R.~Mandal, R.~Sinha and D.~Das,
  Phys.\ Rev.\ D {\bf 90}, no. 9, 096006 (2014)
  [arXiv:1409.3088 [hep-ph]].


\bibitem{Descotes-Genon:2013wba} 
  S.~Descotes-Genon, J.~Matias and J.~Virto,
  Phys.\ Rev.\ D {\bf 88}, no. 7, 074002 (2013)
  [arXiv:1307.5683 [hep-ph]];
  W.~Altmannshofer and D.~M.~Straub,
  Eur.\ Phys.\ J.\ C {\bf 73}, no. 12, 2646 (2013)
  [arXiv:1308.1501 [hep-ph]];
  R.~Gauld, F.~Goertz and U.~Haisch,
  Phys.\ Rev.\ D {\bf 89}, 015005 (2014)
  [arXiv:1308.1959 [hep-ph]].
  A.~J.~Buras and J.~Girrbach,
  JHEP {\bf 1312}, 009 (2013)
  [arXiv:1309.2466 [hep-ph]].
  A.~Datta, M.~Duraisamy and D.~Ghosh,
  Phys.\ Rev.\ D {\bf 89}, no. 7, 071501 (2014)
  [arXiv:1310.1937 [hep-ph]].


\bibitem{Buchalla:1995vs} 
  G.~Buchalla, A.~J.~Buras and M.~E.~Lautenbacher,
  Rev.\ Mod.\ Phys.\  {\bf 68}, 1125 (1996)
  [hep-ph/9512380].



\bibitem{Asatrian:2002va} 
  H.~M.~Asatrian, K.~Bieri, C.~Greub and A.~Hovhannisyan,
  Phys.\ Rev.\ D {\bf 66}, 094013 (2002)
  [hep-ph/0209006].



\bibitem{Ball:2004rg} 
  P.~Ball and R.~Zwicky,
  Phys.\ Rev.\ D {\bf 71}, 014029 (2005)
  [hep-ph/0412079].


\bibitem{Charles:1998dr} 
  J.~Charles, A.~Le Yaouanc, L.~Oliver, O.~Pene and J.~C.~Raynal,
  Phys.\ Rev.\ D {\bf 60}, 014001 (1999)
  [hep-ph/9812358];
  M.~Beneke and T.~Feldmann,
  Nucl.\ Phys.\ B {\bf 592}, 3 (2001)
  [hep-ph/0008255];
  M.~Beneke, T.~Feldmann and D.~Seidel,
  Nucl.\ Phys.\ B {\bf 612}, 25 (2001)
  [hep-ph/0106067].


\bibitem{Beneke:2004dp} 
  M.~Beneke, T.~Feldmann and D.~Seidel,
  Eur.\ Phys.\ J.\ C {\bf 41}, 173 (2005)
  [hep-ph/0412400].


\bibitem{Kruger:1999xa} 
  F.~Kruger, L.~M.~Sehgal, N.~Sinha and R.~Sinha,
  Phys.\ Rev.\ D {\bf 61}, 114028 (2000)
  [Erratum-ibid.\ D {\bf 63}, 019901 (2001)]
  [hep-ph/9907386];
  W.~Altmannshofer, P.~Ball, A.~Bharucha, A.~J.~Buras, D.~M.~Straub and M.~Wick,
  JHEP {\bf 0901}, 019 (2009)
  [arXiv:0811.1214 [hep-ph]];
  J.~Matias, F.~Mescia, M.~Ramon and J.~Virto,
  JHEP {\bf 1204}, 104 (2012)
  [arXiv:1202.4266 [hep-ph]].


\bibitem{Kruger:2005ep} 
  F.~Kruger and J.~Matias,
  Phys.\ Rev.\ D {\bf 71}, 094009 (2005)
  [hep-ph/0502060];
  U.~Egede, T.~Hurth, J.~Matias, M.~Ramon and W.~Reece,
  JHEP {\bf 0811}, 032 (2008)
  [arXiv:0807.2589 [hep-ph]].


\bibitem{Burdman:1998mk} 
  G.~Burdman,
  Phys.\ Rev.\ D {\bf 57}, 4254 (1998)
  [hep-ph/9710550];
  A.~Ali, P.~Ball, L.~T.~Handoko and G.~Hiller,
  Phys.\ Rev.\ D {\bf 61}, 074024 (2000)
  [hep-ph/9910221].



\bibitem{Aaij:2013iag} 
  R.~Aaij {\it et al.}  [LHCb Collaboration],
  JHEP {\bf 1308}, 131 (2013)
  [arXiv:1304.6325, arXiv:1304.6325 [hep-ex]].


\bibitem{Hurth:2013ssa} 
  T.~Hurth and F.~Mahmoudi,
  JHEP {\bf 1404}, 097 (2014)
  [arXiv:1312.5267 [hep-ph]].




\end{thebibliography}

%

\end{document}